\documentstyle[aps,prl,floats,twocolumn,psfig]{revtex}
\begin{document}
\draft

\title{Interaction-induced delocalization of two particles in a random 
potential: Scaling properties}
\author{Felix von Oppen, Tilo Wettig, and Jochen M\"uller}
\address{Max-Planck-Institut f\"ur Kernphysik, 69117 Heidelberg, Germany}
\date{November 7, 1995}

\twocolumn[
\maketitle\widetext\vspace*{-5mm}\leftskip=1.9cm\rightskip=1.9cm
  The localization length   $\xi_2$  for coherent  propagation of  two
  interacting particles in a random potential is studied using a novel
  and  efficient numerical  method. We find   that the enhancement  of
  $\xi_2$ over the  one-particle localization length $\xi_1$ satisfies
  the scaling relation $\xi_2/\xi_1=f(u/\Delta_\xi)$, where $u$ is the
  interaction strength and $\Delta_{\xi}$ the level  spacing of a wire
  of length  $\xi_1$.  The scaling function   $f$ is linear   over the
  investigated parameter range. This   implies that $\xi_2$  increases
  faster with $u$ than  previously  predicted. We  also study a  novel
  mapping of the problem to a banded-random-matrix model.
\begin{abstract}
\\\vspace*{-9mm}
\end{abstract}]

\narrowtext 

While much is known about the  localization properties of one particle
moving in  a  random potential \cite{Kramer},  there are   few secured
results about localization in the presence of interactions between the
particles \cite{Belitz}. In view of the complexity of the interplay of
disorder and interaction, Dorokhov \cite{Dorokhov} and, very recently,
Shepelyansky \cite{Shepelyansky} approached this problem by studying a
simple special  case ---    two  interacting particles  in   a  random
one-dimensional potential --- and  predicted that the interaction  can
lead to   a  significant  delocalization  of   the  pair.  A  possible
realization of this system  are excitons in a disordered semiconductor
\cite{Dorokhov}.    Furthermore,  understanding     the   localization
properties of  two particles in   a random potential  may  lead to new
insights into the role of interactions in the Anderson insulator.

Shepelyansky considered the motion of  two particles interacting by  a
short-range interaction  in   a random potential.  Whenever    the two
particles are  localized  far   apart  compared to   the  one-particle
localization length,  the   effect    of  the  interaction   is   only
exponentially small.  However,  an interesting effect occurs  when the
two particles are localized within about one one-particle localization
length   of  each other.  In  this   case, Shepelyansky constructed an
approximate  mapping of  the problem to  a banded-random-matrix model.
Studying this model numerically, he  predicted that {\it independently
of the statistics of the particles  and the sign of the interaction}\/
there is the possibility of coherent propagation  of the two particles
over distances $\xi_2$ much  larger than the one-particle localization
length $\xi_1$ \cite{Shepelyansky}. He found that
\begin{equation}
   {\xi_2\over\xi_1}\approx{\xi_1\over 32}\left({u\over t}\right)^2,
\label{shepel}
\end{equation}
where  $u$ denotes  the interaction strength   and $t$  is the hopping
matrix element. $\xi_1$ is measured in units  of the lattice constant.
Subsequently,  Imry   \cite{Imry}  has given a   Thouless-type scaling
argument  supporting and generalizing this  result,  and Frahm et al.\
\cite{Frahm}  have   studied    the   problem  numerically    using  a
transfer-matrix  technique, finding a  slower increase of $\xi_2$ with
$\xi_1$  than  predicted  by  Eq.~(\ref{shepel}).  Related results had
been found earlier by Dorokhov \cite{Dorokhov}  for the propagation of
two harmonically bound particles in a random potential.

In this paper, we present a novel and efficient numerical technique to
compute the two-particle  localization length $\xi_2$ directly from  a
microscopic model. This method allows   us to obtain accurate  results
over a wide range of parameters for both bosons and fermions. Our main
result is that  $\xi_2$  obeys   the scaling  relation   $\xi_2/\xi_1=
f(u/\Delta_\xi)$, where  $\Delta_\xi$   is the  single-particle  level
spacing of a  wire  of length $\xi_1$.  We  conjecture,  based  on our
numerical  results, that the exact scaling  function  $f$ is linear at
the center of  the band.  While  our results qualitatively confirm the
prediction   that  a  short-range  interaction  can  lead  to coherent
propagation   of  the pair  over   distances  much   larger than   the
one-particle localization   length,    this  scaling   relation     is
inconsistent with the original  prediction, Eq.~(\ref{shepel}).  It is
an  important   consequence of our results   that   the enhancement of
$\xi_2$ sets in    for  {\it weaker}\/  interactions   than previously
predicted. We also derive and study a novel  mapping of the problem to
a banded-random-matrix model.   A  combination  of scaling with   this
banded-random-matrix model suggests the  possibility that the validity
of our principal results extends to quasi-one-dimensional wires.

Our starting point is the  Anderson Hamiltonian $H_0$ for two spinless
particles in a   one-dimensional random potential with   an additional
Hubbard-type interaction $U$,
\begin{eqnarray}
  H&=&t\sum_{n,m}\left\{|n,m\rangle\langle n\!+\!1,m|+|n,m\rangle\langle 
      n,m\!+\!1|+{\rm h.c.}\right\}\nonumber\\
      & & \mbox{} +\sum_{n,m}|n,m\rangle(V_n+V_m)\langle n,m|+U.
\label{model}
\end{eqnarray}
The random  site energies $V_n$ are drawn  uniformly from the interval
$[-W/2,W/2]$. The hopping  matrix element $t$ will  be set to unity in
the following.   We   parameterize the disorder  by   the one-particle
localization length\cite{Kramer} $\xi_1\!=\!105(t/W)^2$ in the absence
of the interaction.  For bosons, we choose an on-site interaction with
matrix     elements  $\langle   n,m|U|n^\prime,m^\prime\rangle\!=\!u\,
\delta_{n,n^\prime}\, \delta_{m,m^\prime}\delta_{n,m}$,  for  spinless
fermions a  nearest-neighbor interaction with matrix elements $\langle
n,m|U|n^\prime,m^\prime\rangle\!=\!u[\delta_{n,m+1}+\delta_{n,m-1}]
\delta_{n,n^\prime} \delta_{m,m^\prime}$.   In  the    following   our
numerical method is described for bosons. The extension to fermions is
straightforward.  To study the two-particle localization properties of
the Hamiltonian (\ref{model}) we focus  on the matrix elements of  the
two-particle Green function
\begin{equation}
  G=(E-H_0-U)^{-1}
\end{equation}  
between   doubly-occupied  sites   $|n,n\rangle$.   We   define    the
two-particle localization length $\xi_2$ for coherent transport by the
exponential decrease with distance of these matrix elements,
\begin{equation}
  {1\over\xi_2}=-\lim_{|n-m|\to\infty}
              {1\over|n-m|}\ln|\langle n,n|G|m,m\rangle|.
\label{2ploc}
\end{equation}
This   identification   is  certainly   reasonable  as    long  as the
doubly-occupied sites  do not effectively  decouple from the remaining
Hilbert space because    of the interaction,   i.e.,  as long  as  the
interaction strength  $u$  is smaller than or    of the order   of the
hopping  matrix  element  $t$. Our   numerical   approach exploits the
observation that, for the interaction  $U$,  a closed equation can  be
derived for  these matrix elements. This reduces  the dimension of the
relevant Hilbert space  from  $N^2$ to  $N$ (with  $N$ the number   of
sites) which enables us  to study systems with up  to 1000 sites.  The
Dyson equation for the two-particle Green function is
\begin{equation}
           G=G_0+G_0UG,
\end{equation}
where $G_0=(E-H_0)^{-1}$ denotes the  Green function in the absence of
the interaction. The interaction can  be written as $U\!=\!uP$,  where
$P$   denotes    the   projector  onto  the    doubly-occupied  sites,
$P|n,m\rangle\!=\!  \delta_{n,m}|n,m\rangle$. Hence, multiplying   the
Dyson  equation by  $P$  on both sides  and  writing $U\!=\!uP^2$, one
obtains a  closed equation  for    the  two-particle Green    function
projected onto doubly-occupied sites,
\begin{equation}
  {\tilde G}={\tilde G}_0+u{\tilde G}_0{\tilde G}.
\end{equation}
Here we defined ${\tilde G}\!=\!PGP$ and ${\tilde
G}_0\!=\!PG_0P$. Solving this equation for ${\tilde G}$ one has
\begin{equation}
    {\tilde G}={{\tilde G_0}\over u}{1\over{1/u-{\tilde G}_0}}.
\label{proGreen}    
\end{equation}
In the site basis, the unperturbed Green function  ${\tilde G}_0$ is a
banded   matrix whose   matrix elements   decrease  exponentially with
distance  on the scale  $\xi_1/2$. Therefore, we compute $\xi_2$ using
only  the second factor in (\ref{proGreen})  from which any long-range
behavior of $G$ must arise. This is very useful for numerical purposes
because this  factor can be interpreted  as the  Green function of the
``Hamiltonian'' ${\tilde  G}_0$ at ``energy''  $1/u$.  This enables us
to employ   the efficient recursive  Green-function  method for banded
Hamiltonian  matrices   \cite{Huckestein}   to find  the  two-particle
localization length. We obtain the exact ${\tilde G}_0$ \cite{foot0},
\begin{equation}
  \langle n,n|{\tilde G}_0(E)|m,m\rangle=\sum_{i,j}{\phi_i(n)\phi_j(n)
    \phi_i^*(m)\phi^*_j(m)\over{E-E_i-E_j}},
\label{2pgreen} 
\end{equation}
by solving the Anderson model in the absence  of the interaction. Here
the $\phi_i$ are the  exact  single-particle wave functions.  Clearly,
our method of computing $\xi_2$  is accurate whenever the  enhancement
factor $\xi_2/\xi_1$ is sufficiently large.  Deviations from the exact
$\xi_2$ arise for small $u$ and small $\xi_1$ where the enhancement is
weak. For small $u$ this can be  easily seen because the second factor
in (\ref{proGreen}) gives $\lim_{u\to0}\xi_2=0$, while the exact limit
is $\xi_1/2$.

We have studied the  two-particle localization length $\xi_2$ for both
fermions   and   bosons    for  one-particle   localization    lengths
$4.2\leq\xi_1\leq105$  and interaction  strengths $0\leq  u\leq1$.  We
find  that it is  sufficient  to  use systems  with $N\!=\!500$  sites
except  for  the   two largest  values   of   $\xi_1$ where   we  used
$N\!=\!1000$.  For    each value  of   $\xi_1$  we averaged    over 50
realizations of the disorder. In  Fig.~1 we have plotted $\xi_2/\xi_1$
as a function of $u\xi_1/t$ at the center of  the band (E=0).  We have
included data for ten values of $u$ for each  of five values \cite{w1}
of $\xi_1$.  The observed scaling behavior
\begin{equation}
    {\xi_2\over\xi_1}={\tilde f}(u\xi_1/t) 
\label{scaling}
\end{equation}
is the central result of this paper. While the data  in Fig.~1 are for
$E=0$, we find that the same scaling behavior holds also away from the
center of the band \cite{Oppen}.

\begin{figure}
\centerline{\psfig{figure=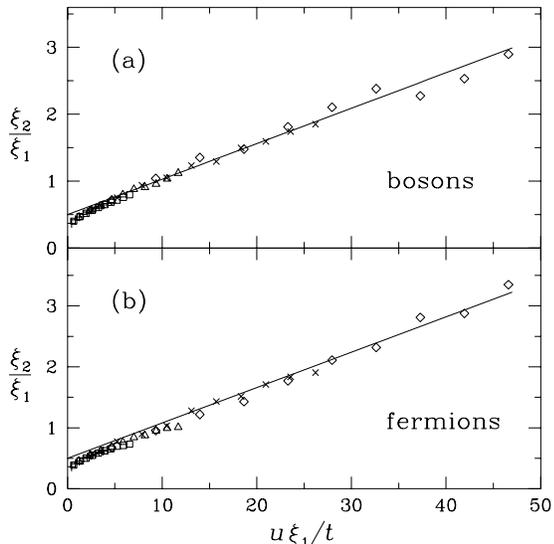,width=75mm}}
\caption{Scaling plot $\xi_2/\xi_1={\tilde f}(u\xi_1/t)$ for the
  two-particle localization  length   $\xi_2$ of (a)  bosons   and (b)
  fermions as a function of  interaction strength $u$ and one-particle
  localization length $\xi_1$. Ten values of $u$ are included for each
  of  the five  values of disorder   $W=5$ (pluses),  $W=4$ (squares),
  $W=3$ (triangles), $W=2$ (crosses), and $W=1.5$ (diamonds). The full
  lines show    that    the linear  behavior    for   large $u\xi_1/t$
  extrapolates to $\xi_2/\xi_1=1/2$   for $u\to0$. The deviation  from
  linear behavior for  small $u\xi_1/t$ is  most likely an artifact of
  our numerical method.}
\end{figure}

This scaling behavior implies    that the scale for  the   interaction
strength $u$  is the energy $t/\xi_1$  which can be interpreted as the
single-particle level spacing $\Delta_\xi=\pi  t/\xi_1$  of a wire  of
length $\xi_1$.  The scaling function  in Fig.~1 can be  fit well by a
straight line for   sufficiently large $u/\Delta_\xi$ while there  are
deviations from  linear  behavior  for  small values of    the scaling
variable. It  is natural to suppose that  these deviations from linear
behavior  are     an artifact of  disregarding    the  first factor in
Eq.~(\ref{proGreen})  in computing $\xi_2$.  In fact,  as shown by the
full line  in  Fig.~1, the linear  behavior  for  large $u/\Delta_\xi$
extrapolates  to  $\xi_2/\xi_1=1/2$ for $u/\Delta_\xi\to0$ as expected
for the exact  two-particle localization length  $\xi_2$ computed from
the full expression for  ${\tilde G}$. Hence,  one may conjecture that
the exact scaling relation at the center of the band has the form
\begin{equation}
     {\xi_2\over\xi_1}={1\over2}+C{|u|\over\Delta_\xi},
\label{linear}
\end{equation}
where  $C\approx0.17$  for  bosons   and  $C\approx0.18$  for fermions
\cite{comp}.  We used the fact that at the center of the band there is
an   exact symmetry between  attractive  and repulsive interactions so
that  $\xi_2$ depends only on  the  absolute value  of  $u$. Note that
while this result is presumably valid for arbitrary values of $\xi_1$,
one  expects the  scaling relation to  break down  for large $u$ where
$u/t$ should become   a  relevant parameter due   to density-of-states
effects.

Shepelyansky's  original  prediction, Eq.~(\ref{shepel}), is {\it not}
consistent  with the scaling relation   (\ref{scaling}).  E.g., at the
center of   the band  our   results show that  while   $\xi_2$ depends
quadratically on    $\xi_1$ as previously   predicted,  it exhibits an
unexpected linear (instead of quadratic) dependence on the interaction
strength  $u$.  More generally,  the  scaling  (\ref{scaling}) implies
that   the enhancement effect sets in   for weaker interactions $u\sim
t/\xi_1$   [compared    to   $u\sim   t/\xi_1^{1/2}$    according   to
Eq.~(\ref{shepel})]   than   originally  predicted.  This   result  is
surprising   in view  of  the following  estimate.   It  may be argued
\cite{Imry}  that $\xi_2$ should      deviate from $\xi_1$    once the
two-particle product  states $|\phi_i,\phi_j\rangle$ of  (unperturbed)
energy $E_{i,j}$ are strongly  mixed by the interaction $U$. According
to      perturbation    theory,     strong     mixing    occurs   when
$\langle\phi_1,\phi_2|U|\phi_3, \phi_4\rangle/(E_{1,2}-E_{3,4})$ is of
order  unity.   Each  $|\phi_i,\phi_j\rangle$   is  typically  coupled
appreciably to $\xi_1^2$ states.  The corresponding matrix element can
be  estimated  \cite{Shepelyansky,Imry}  as  $u/\xi_1^{3/2}$   and the
energy denominator as  $t/\xi_1^2$.  Thus, according to this estimate,
strong mixing    occurs   once $u\sim    t/\xi_1^{1/2}$.   Since  this
interaction strength is  large compared to $\Delta_\xi$,  a comparison
with our result would suggest that, surprisingly, strong mixing of the
two-particle product states is   not necessary for the  enhancement of
the two-particle localization length.

Originally, Shepelyansky \cite{Shepelyansky} approached the problem by
an  approximate mapping to a banded-random-matrix  model. We have also
investigated an alternative random-matrix model  which is suggested by
Eq.~(\ref{proGreen})  due to  the  band-matrix  structure  of ${\tilde
G}_0$. An extension of this banded-random-matrix model will be applied
below to study two interacting   particles in a  quasi-one-dimensional
wire. In contrast to ordinary banded random  matrices we find that the
matrix elements $g$ of ${\tilde G}_0(E=0)$ have a Cauchy distribution,
$P(g)\!=\!(\Gamma/\pi)/(\Gamma^2+g^2)$.  To obtain this   distribution
function we argue as  follows.  For definiteness, consider a  diagonal
matrix element of ${\tilde G}_0$.  Due to the  localized nature of the
wave functions there are of the order of $\xi_1^2$ terms in the sum in
(\ref{2pgreen}). Furthermore,  normalization   implies  that the  wave
functions are  of  order  $1/\sqrt{\xi_1}$  within a  region  of  size
$\xi_1$.  Hence,  $\langle n,n|{\tilde G}_0|n,n \rangle\sim(1/\xi_1^2)
\sum_{k=1}^{\xi_1^2} (1/x_k)$, where $x_k\!\sim\!-\!E_i\!-\!E_j$ is  a
random variable  in the  range   $-4t\,{{\raise-3pt\hbox{$\scriptstyle
<$}}\atop{\raise4pt\hbox
{$\scriptstyle\sim$}}}\,x_k\,{{\raise-3pt\hbox{$\scriptstyle <$}}\atop
{\raise4pt\hbox{$\scriptstyle\sim$}}}\,4t$.  Thus, the matrix elements
of the Green function  are  given by  averages over random   variables
whose second moments   diverge. Neglecting  correlations  between  the
$x_k$,  the   central-limit theorem   implies for   sufficiently large
$\xi_1$  that the diagonal matrix elements  have a Cauchy distribution
of  width $\Gamma\sim 1/t$  \cite{Bouchaud}.  The same argument can be
made   for the off-diagonal matrix  elements.  Their width $\Gamma$ is
reduced  by a factor of   $\exp(-2|n-m|/\xi_1)$.  As shown in  Fig.~2,
these conclusions are well supported by numerical results.

\begin{figure}
\centerline{\psfig{figure=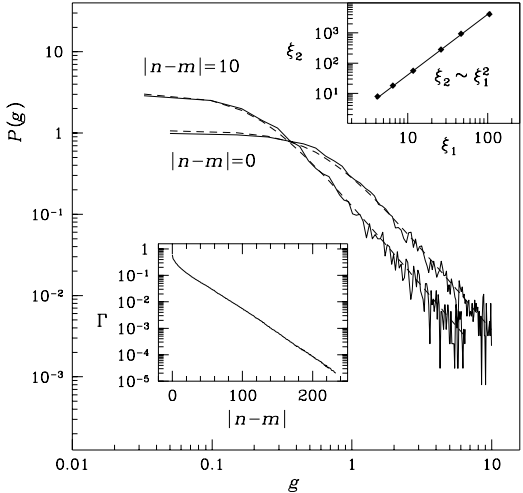,width=75mm}}
\caption{Distributions of diagonal and off-diagonal matrix elements
  $g$ of the projected two-particle  Green function ${\tilde G}_0$ for
  $\xi_1\!=\!46.6$ (solid lines).  The dashed lines are fits to Cauchy
  distributions.  Their  width  $\Gamma$ decreases  exponentially away
  from  the diagonal on  the  scale $\xi_1/2$   as shown in  the lower
  inset. The upper inset shows $\xi_2$ vs $\xi_1$ as obtained from the
  banded-random-matrix   model discussed   in  the text.    These data
  obtained for systems with   $10^6$ sites confirm our  arguments that
  $\xi_2\sim\xi_1^2$ in this model.}
\end{figure}

We argue  that,     for   sufficiently large  $u$,    the    resulting
banded-random-matrix   model    predicts    $\xi_2\!\sim\!\xi_1^2$  in
agreement with Eq.~(\ref{shepel}) when neglecting correlations between
the  matrix  elements.   This  result  would follow  immediately  from
analytical results  for banded random matrices  if the distribution of
the  matrix  elements had  a    finite variance.  In this   case,  the
localization  length  is proportional to the   square of the bandwidth
\cite{Fyodorov}. The same result holds true for banded Cauchy matrices
for the  following reason: Since the eigenstates  of the banded Cauchy
matrix are localized, they effectively sample only  a finite number of
matrix  elements  drawn  from  the Cauchy  distribution.  Hence, there
exists a corresponding  typical  largest matrix element  $g_{\rm max}$
\cite{Bouchaud}. Beyond $g_{\rm  max}$ the Cauchy distribution can  be
cut off, and the resulting  effective distribution of matrix  elements
has a finite variance. This implies that banded Cauchy matrices belong
to the same universality class  as ordinary banded random matrices. We
confirmed this  conclusion numerically  by computing  the localization
length of  banded Cauchy matrices as  a function of bandwidth as shown
in the upper inset of Fig.~2. We have  also studied the $u$ dependence
of $\xi_2$ predicted  by this banded-random-matrix model. However,  we
find that the $\xi_2$ computed from this  model does {\it not} exhibit
the scaling (\ref{scaling}) found for  the exact solution. Presumably,
this is due  to correlations between  the matrix  elements of ${\tilde
G}_0$.  For example,  correlations   in the  exact  ${\tilde G}_0$ are
implied  by   Eq.~(\ref{2pgreen})   for   exceptionally  large  matrix
elements. Large matrix elements are due  to small energy denominators.
Each product   state with energy close  to  $E$ leads to  a correlated
$\xi_1\times\xi_1$ block of  large matrix elements  in ${\tilde G}_0$.
Such correlations are neglected in the banded-random-matrix model.

It is  an interesting  problem  to study two-particle  localization in
more  than one dimension.  In the  absence of  an understanding of the
physical  origin of  the scaling parameter  $u/\Delta_\xi$,  it is not
clear  how  to  generalize   our   results  to   these  cases.     For
quasi-one-dimensional wires with a finite number  of channels $M$, one
easily derives a generalized banded-random-matrix-model. Assuming that
this  banded-random-matrix   model   again  correctly  predicts    the
dependence of $\xi_2$ on the  bandwidth and combining the result  with
scaling suggests that  the  scaling   function remains linear.     For
quasi-one-dimensional wires  we can  order the  doubly-occupied  sites
sequentially along the longitudinal direction.  When $\xi_1$ is larger
than the transverse dimensions of the wire,  the bandwidth of ${\tilde
G}_0$  is equal to   $M\xi_1$ yielding a  corresponding ``localization
length''  $(M\xi_1)^2$.  The   actual   localization   length in   the
longitudinal  direction is smaller by  a  factor $M$, hence $\xi_2\sim
M\xi_1^2$.  Finally, assuming the  above  scaling behavior and  noting
that $\Delta_\xi\sim t/M\xi_1$,  we obtain a linear  scaling function,
$\xi_2/\xi_1\sim|u|/\Delta_\xi$.

In summary, we  have studied the interaction-induced delocalization of
two particles  in a  one-dimensional  random potential by a  novel and
efficient numerical approach.   We have  found that the   two-particle
localization   length $\xi_2$   for coherent propagation   of the  two
particles satisfies the scaling relation $\xi_2/\xi_1=f(u/\Delta_\xi)$
as  a    function   of interaction  strength    $u$   and one-particle
localization length $\xi_1$. This implies that the  effect sets in for
weaker interactions than  previously predicted.  At  the center of the
band  our data suggest  that  the   scaling  function is linear.    At
present,  we   do not have    a good  physical   understanding of this
unexpected  scaling behavior.  It  will be  interesting to see whether
the  scaling   found in  this  paper   can  be  generalized  to higher
dimensions or whether it is a specific feature of one dimension.

It would  also    be interesting to  study   implications  of coherent
propagation due to   interactions  at  finite particle density.    Our
numerical  approach can  be  extended   to study  the  propagation  of
quasiparticle pairs in  the    Anderson insulator. This will   be  the
subject of a separate publication.

One of us (FvO) is particularly indebted to  B.\ Alt\-shu\-ler and A.\
Luther for important discussions which eventually led to a significant
revision of a previous version of the  manuscript. We also acknowledge
useful  discussions   with  G.\  Hackenbroich,    B.\ Huckestein,  A.\
M\"uller-Groeling, D.L.\ Shepelyansky, and H.A.\ Weidenm\"uller.


\begin{references}

\bibitem{Kramer} For a review, see B.\ Kramer and A.\ MacKinnon, Rep.\ 
  Prog.\ Phys.\ {\bf 56}, 1469 (1993).

\bibitem{Belitz} For a review, see D.\ Belitz and T.R.\ Kirkpatrick,
  Rev.\ Mod.\ Phys.\ {\bf 66}, 261 (1994).

\bibitem{Dorokhov} O.N.\ Dorokhov, Zh.\ Eksp.\ Teor.\ Fiz.\ {\bf 98}, 
  646 (1990) [Sov.\ Phys.\ JETP {\bf 71}, 360 (1990)].

\bibitem{Shepelyansky}  D.L.\  Shepelyansky, Phys.\  Rev.\ Lett.\ {\bf
    73}, 2607 (1994);  F.\ Borgonovi and  D.L.\ Shepelyansky, preprint
  (1995).

\bibitem{Imry} Y.\ Imry, Europhys.\ Lett.\ {\bf 30}, 405 (1995).

\bibitem{Frahm} K.\ Frahm, A.\ M\"uller-Groeling, J.-L.\ Pichard, and
  D.\ Weinmann, Europhys.\ Lett.\  {\bf 31}, 169 (1995); D.\ Weinmann,
  A.\ M\"uller-Groeling,  J.-L.\  Pichard,  and  K.\  Frahm,  preprint
  (1995).

\bibitem{Huckestein} B.\ Huckestein, Rev.\ Mod.\
  Phys.\ {\bf 67}, 357 (1995).

\bibitem{foot0} Note that this Green function is properly symmetrized
  although this may not be apparent because of the projection onto
  doubly-occupied sites.

\bibitem{w1} We have not included our data for $W=1$ ($\xi_1=105$) in
  Fig.~1.  While they are consistent   with a linear scaling  function
  $f$,  the corresponding error bars are  rather large because $\xi_2$
  is comparable to the system size.

\bibitem{Oppen} F.\ von Oppen and T.\ Wettig, unpublished.

\bibitem{comp} Note that the dependence of $\xi_2$ on $\xi_1$ involves
  both  a linear and a  quadratic term. From a double-logarithmic plot
  of  $\xi_2$ vs  $\xi_1$  over   a  limited parameter  range,    this
  dependence  was identified in Ref.~\cite{Frahm} as  a power law with
  non-integer exponent between one and two.

\bibitem{Bouchaud} E.g., see J.-P.\ Bouchaud and A.\ Georges, Phys.\ 
  Rep.\ {\bf 195}, 127 (1990).

\bibitem{Fyodorov} Y.V.\ Fyodorov and A.D.\ Mirlin, Phys.\ Rev.\
  Lett.\ {\bf 67}, 2405 (1991).

\end{references}
\end{document}